\documentclass[12pt]{iopart}
\usepackage[utf8]{inputenc}
\usepackage{lipsum}
\usepackage{braket}
\usepackage{cite}
\usepackage{amssymb}
\usepackage{upgreek}
\usepackage{outlines}
\usepackage{graphicx}
\usepackage{placeins}
\usepackage{bbold}
\usepackage[normalem]{ulem}
\usepackage{dsfont}
\usepackage{hyperref}
\usepackage{xcolor}
\hypersetup{colorlinks=true, citecolor=blue, urlcolor=blue, linkcolor=blue}
\usepackage[shortlabels]{enumitem}

\newcommand{\ku}{\ket{\uparrow}}
\newcommand{\kd}{\ket{\downarrow}}

\newcommand{\ww}{\textsl{w}}
\newcommand{\sww}{\mbox{\scriptsize$\ww\!\!\:$\normalsize}}
\newcommand{\etaeff}{\tilde{\eta}}

\begin{document}
\title{Tunable transverse spin-motion coupling for quantum information processing}
\author{Adam D. West, Randall Putnam, Wesley C. Campbell, Paul Hamilton}
\address{UCLA, Department of Physics and Astronomy, 475 Portola Plaza, Los Angeles, California 90095, USA}
\ead{adam@physics.ucla.edu}

\date{\today}

\begin{abstract}
Laser-controlled entanglement between atomic qubits (`spins') and collective motion in trapped ion Coulomb crystals requires conditional momentum transfer from the laser. Since the spin-dependent force is derived from a spatial gradient in the spin-light interaction, this force is typically longitudinal --- parallel and proportional to the average laser $k$-vector (or two beams' $k$-vector difference), which constrains both the direction and relative magnitude of the accessible spin-motion coupling. Here, we show how momentum can also be transferred perpendicular to a single laser beam due to the gradient in its transverse profile.
By controlling the transverse gradient at the position of the ion through beam shaping, the relative strength of the sidebands and carrier can be tuned to optimize the desired interaction and suppress undesired, off-resonant effects that can degrade gate fidelity.  We also discuss how this effect may already be playing an unappreciated role in recent experiments.

\end{abstract}
\submitto{Quantum Science and Technology}

\section{Introduction}
Quantum computers based on trapped atomic ions use entanglement between the atomic qubits and collective motion to mediate conditional quantum logic between spatially separated qubits \cite{Bruzewicz2019}. This spin-motion entanglement is produced by applying a spatially-varying interaction with an electromagnetic field that gives a spin-dependent force. In laser-driven, ion-ion entangling gates, this force is derived from the longitudinal gradient of the electric field of a laser beam (or, for Raman processes, a pair of beams), in which case the direction of spin-motion coupling is fixed by the laser beam propagation axes \cite{Leibfried2003}. This precludes direct control of ion motion perpendicular to the beam, and also fixes the relative strengths of the resonant spin-only and spin-motion couplings. In many experiments using surface electrode traps, optical access is restricted to be parallel to the surface plane \cite{Romaszko2020Engineering}; this restriction makes it difficult to access motion perpendicular to the plane, both for cooling and coherent operations.

Two workarounds to access out of plane motion are the development of traps with tilted principal axes  \cite{Seidelin2006,Wesenberg2008,Chuang2008,Allcock2010,Stick2010}, or the introduction of time-dependent cross-coupling potentials \cite{Gorman2014}.  These indirect techniques take advantage of the approximate separability of the secular motion into components along the principal axes of the trap to provide access to part of the motion (the secular component), but direct access to the full motional state (for instance, to diagnose excess micromotion) remains challenging \cite{Allcock2010,Narayanan2011Electric}.  Alternative approaches for controlling spin-motion coupling using static and near-field gradients are being pursued by some groups \cite{Ospelkaus2008,Ospelkaus2011,Timoney2011,Harty2016,Sutherland2019,Zarantonello2019,Srinivas2019,Sutherland2020}, but are also constrained by the fixed electrode geometry.

Here, we show that the transverse, as opposed to longitudinal, gradient of the spin-light interaction can also be used to produce and control spin-motion entanglement, even perpendicular to the laser propagation direction.
By adjusting the spatial profile and/or position of the beam, the strength of motional sidebands can be tuned, even to the point where the carrier transition is fully suppressed.  By extinguishing the carrier during sideband operations and extinguishing the sidebands during carrier operations, this flexibility has the potential to suppress errors from off-resonant transitions \cite{Ozeri2007}.
%This has the potential to significantly reduce gate errors associated with unwanted photon scattering events \cite{Ozeri2007}.
As a proof of principle, we demonstrate this transverse spin-motion coupling using a single trapped ion. The stimulated Raman spectrum driven in a co-propagating beam geometry shows motional sidebands driven by the beam's transverse intensity gradient, and we show that their strength can be tuned by varying the ion temperature, in agreement with the model.

\section{Theory}
We consider a laser-driven electronic transition in a single trapped ion and show how the finite transverse extent of the beam can change the motional state perpendicular to the beam, even when the (conventional longitudinal) Lamb-Dicke factor is essentially zero. Since the technique presented here is applicable to every type of electronic transition used for quantum information processing (E2, E3, stimulated Raman, etc.), we present it without reference to the details of the internal state manipulation where possible and point out where differences may arise. We assume that the wavevector of the laser field (or wavevector difference, for stimulated Raman transitions) is aligned with $+\mathbf{\hat{z}}$, which we also assume is a principal axis of the trapping potential such that the longitudinal gradient cannot couple to motion in the $x$-$y$ plane. For simplicity, we consider motion along only the $x$ direction and neglect the other two; a full treatment that includes $y$ and $z$ can be constructed in a straightforward manner. We can write the matrix element associated with this transition as
\begin{equation}
    \Omega_{n^\prime,n} = \Omega_0\braket{n'|f(\ww,x)|n}
\end{equation}
where $n$ ($n'$) is the initial (final) motional state along $x$ and the function $f(\ww,x)$ is the transverse spatial profile of the laser-ion coupling, $\Omega(x) \equiv \Omega_0 f(\ww,x)$. We absorb all of the electronic transition details in $\Omega_0$ and assume the atomic matrix element is proportional to $f$\footnote{Recent work has examined how a transverse electric field profile can drive an electronic, rather than motional, transition \cite{Schmiegelow2016Transfer}.}.

While the beam profile can in principle have a variety of functional forms we will first assume it is Gaussian, with $f(\ww,x) \equiv \exp(-2x^2/\ww^2)$ (shortly, we also consider the case of a TEM$_{10}$ mode). In the case of a stimulated Raman transition, $f(\ww,x)$ is the product of the electric fields $E_1 E_2^*$ and $\ww = w_0$ (the waist of the Gaussian beam, $w_0$, defined as the $1/e^2$ intensity radius); in the case of a single photon transition (e.g.\ E2 or E3), $f(\ww,x)$ is the profile of the electric field and $\ww = \sqrt{2} w_0$.

\begin{figure}[!ht]
    \centering
    \includegraphics[width=\linewidth]{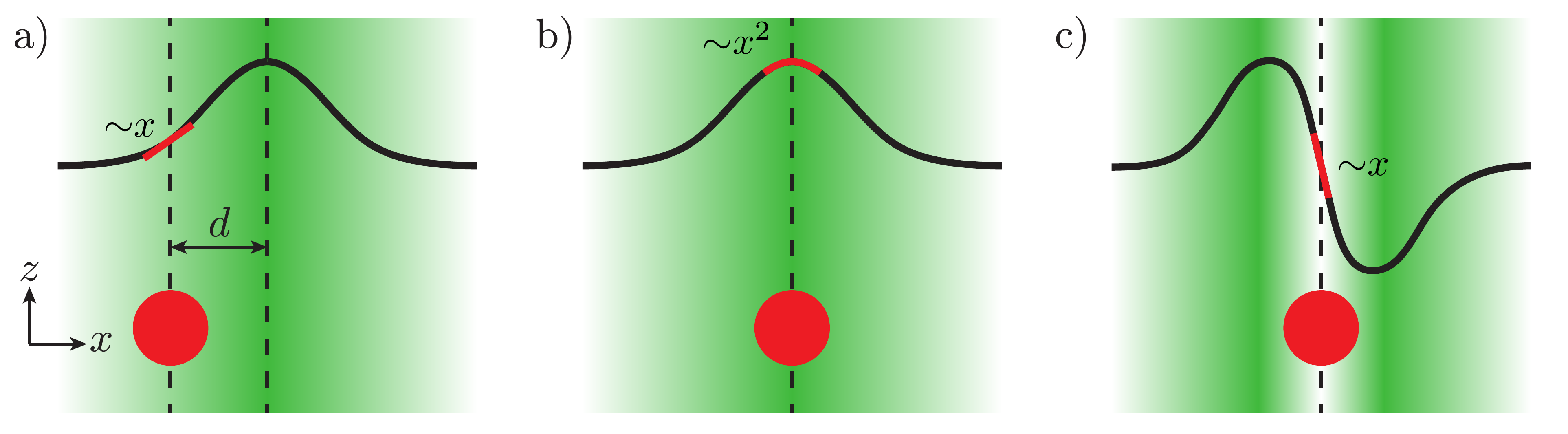}
    \caption{Schematic showing the geometry considered. A laser beam directed along $z$ is incident on a trapped ion (red). a) and b) show the case where the interaction strength has a Gaussian (TEM$_{00}$) transverse profile (black solid line). Depending on the beam position, the profile at the ion can be approximately linear or quadratic (red solid line), coupling to first- or second-order sidebands, respectively. c) shows the case where the profile is produced by a TEM$_{10}$ mode, which suppresses carrier transitions while still coupling to motion.}
    \label{fig:schematic}
\end{figure}

We will treat the spatial profile of the beam(s) by Taylor expanding about the ion's equilibrium position ($x\!=\!0$) to second order in $x$. A Gaussian spatial profile that is offset from the ion's equilibrium position by a distance $d$ (that is, $f(\ww, x-d)$, as shown in fig.~\ref{fig:schematic}a)) produces matrix elements of the following form (up to second order in $x_0$):
% \begin{widetext}
% \begin{align}
%     \Omega_{2\gamma} &= \Omega_{2\gamma}^0e^{-d^2/w^2}\left[\left(1+\left(\frac{8d^2}{w^4}-\frac{2}{w}\right)x_0^2(2n+1)\right)\braket{n'|n} + \frac{4dx_0}{w^2}\left(\sqrt{n}\braket{n'|n-1} + \sqrt{n+1}\braket{n'|n+1}\right) +\right.\nonumber\\ &\hspace{67pt}\left.\left(\frac{8d^2}{w^4}-\frac{2}{w^2}\right)x_0^2\left(\sqrt{(n-1)n}\braket{n'|n-2} + \sqrt{(n+1)(n+2)}\braket{n'|n+2}\right)\right],\label{eq:W2disp}
% \end{align}
% \end{widetext}
% \begin{equation}
%     \hspace{50pt}\fl\Omega_{2\gamma}\propto
%     \left\{
%     \begin{array}{cc}
%     \left(1+\left(\frac{8d^2}{w^4}-\frac{2}{w^2}\right)x_0^2(2n+1)\right) & \quad n'=n\\
%     \frac{4dx_0}{w^2}\sqrt{n} & \quad n'=n-1\\
%     \frac{4dx_0}{w^2}\sqrt{n+1} & \quad n'=n+1\\
%     \left(\frac{8d^2}{w^4}-\frac{2}{w^2}\right)x_0^2\sqrt{(n-1)n} & \quad n'=n-2\\    
%     \left(\frac{8d^2}{w^4}-\frac{2}{w^2}\right)x_0^2\sqrt{(n+1)(n+2)} & \quad n'=n+2
%     \end{array}
%     \right.\label{eq:W2disp}
% \end{equation}
\begin{equation}
    \hspace{50pt}\fl\Omega_{n',n}= \Omega_0\, f(\ww,-d) \, \sqrt{\frac{n_>!}{n_<!}} \times
    \left\{
    \begin{array}{cc}
    1+ \frac{4x_0^2}{\sww^2} \left( \frac{4d^2}{\sww^2} - 1 \right)(n+\frac{1}{2}) & \quad \Delta n = 0\\
    \frac{4dx_0}{\sww^2} & \quad |\Delta n| = 1\\
    \frac{2x_0^2}{\sww^2} \left( \frac{4d^2}{\sww^2} - 1 \right) & \quad |\Delta n| = 2
    \end{array},
    \right.\label{eq:SidebandStrengths}
\end{equation}
where $n_<$ ($n_>$) is the lesser (greater) of $n$ and $n'$, and $x_0 \equiv\sqrt{\hbar/2m\omega}$ is the motional mode's ground state wavefunction size, with $m$ the mass and $\omega$ the secular frequency. We note that when $d=0$ the Rabi frequency of all odd-order sidebands vanishes, as can be seen in Eq.~\ref{eq:SidebandStrengths} for $|\Delta n|=1$. In fact, when $d=0$, exact expressions for $\Omega_{n',n}$ can be obtained, and are provided in \ref{sec:rabi_exact}.

In the Lamb-Dicke regime, a simple analytic expression describes the longitudinal spin-motion coupling, to lowest order in the Lamb-Dicke parameter $\eta\equiv kx_0$ (for wavevector $k$) \cite{Wineland1998}. We compare this to the case of transverse spin-motion coupling by defining an effective Lamb-Dicke parameter, $\etaeff_{(s)}$, where $s$ is the sideband order. As an example, if $d=\ww/2$, for the first order sidebands we have
\begin{equation}
    \Omega_{n^\prime,n} = \etaeff_{(1)} \,\Omega_0 f(\ww,-\ww/2)\sqrt{\frac{n_>!}{n_<!}}
\end{equation}
with
\begin{equation}
    \etaeff_{(1)} \equiv \frac{2x_0}{\ww} \approx 0.014\sqrt{\frac{100~{\rm amu}}{m}}\sqrt{\frac{2\pi\times1~{\rm MHz}}{\omega}}\frac{1~\upmu{\rm m}}{\ww}.\label{eq:etaformula}
\end{equation}
This same expression, Eq.~(\ref{eq:etaformula}), is applicable to the second sidebands (with $\Omega_{n^\prime,n} \propto \etaeff_{(2)}^2/2$) when $d=0$. Unlike longitudinal spin-motion coupling from a plane wave, where the $p$th order sideband term is approximately proportional to $\eta^p/(p!)$, the expressions for the sideband strengths from transverse coupling are a function of the beam profile and position, and should be calculated individually for each sideband order.

The intuitive conclusion that we can draw is that transverse coupling to the ion motion is
%could also say it becomes significant for large d
significant once the wavefunction size, $\sqrt{n}x_0$, becomes comparable to the transverse profile size, $\ww$. As the spatial extent of the beam becomes smaller, the corresponding momentum spread increases, in accorance with the uncertainty principle. For the case of a stimulated Raman transition using a single focussed beam, one can associate an effective wavevector, $k_{\rm eff}$, with this momentum spread. An equivalent value of $k_{\rm eff}$ can be achieved with two infinite plane waves crossing at an angle equal to the half-cone divergence angle of the single beam, $\theta \equiv \frac{\lambda}{\pi w_0}$, i.e.~the coupling strength with the single focussed beam is half that for a pair of crossed plane waves. This is a spatial manifestation of Ramsey's famous factor of two \cite{KleppnerRamsey}.

Having seen that transverse coupling to odd-order sidebands disappears with a centered TEM$_{00}$ beam, we now show that coupling to even-order sidebands (and carrier) can be extinguished if $f(\ww,x)$ is an odd function of $x$, such as with a TEM$_{10}$ mode (cf.\ fig.~\ref{fig:schematic}c)) driving either a single photon transition (such as E2) or one of the arms of a stimulated Raman transition (with the other arm uniform intensity). Here, the Rabi frequency, $\Omega(x) \equiv \Omega_0 f(\ww,x)$, vanishes at the equilibrium position of the ion and has odd parity. For a TEM$_{10}$ beam with waist $w_0$, this configuration produces the same Rabi coupling for the single-photon and Raman cases, $f(w_0,x-d) \equiv H_1(\sqrt{2}(x-d)/w_0)\exp(-(x-d)^2/w_0^2)= 2\sqrt{2}\,\frac{x-d}{w_0}\exp(-(x-d)^2/w_0^2)$ where $H_1(x)$ is the first Hermite polynomial. Once again expanding to second order in $x_0$ gives the matrix elements for the carrier and the first and second sidebands:
% \begin{eqnarray}
%     f(w,x-d) &= 2\sqrt{2}\frac{x-d}{w}e^{-(x-d)^2/w^2}\\
%     &= 2\sqrt{2}e^{-d^2/w^2}\left[-\frac{d}{w}+\frac{x(w^2-2d^2)}{w^3}+\frac{x^2(3dw^2-2d^3)}{w^5}+\cdots\right].
% \end{eqnarray}
% \textcolor{red}{The matrix element is thus}
% \begin{equation}
%     \braket{n'|f(w,x-d)|n} = \sqrt{\frac{n_>!}{n_<!}}\times
%     \left\{
%     \begin{array}{cc}
%     -2\sqrt{2}e^{-d^2/w^2}\frac{d}{w}+\mathcal{O}(x_0^2) & \quad \Delta n = 0\\
%     2\sqrt{2}e^{-d^2/w^2}\frac{x_0(w^2-2d^2)}{w^3} & \quad |\Delta n| = 1\\
%     2\sqrt{2}e^{-d^2/w^2}\frac{x_0^2(3dw^2-2d^3)}{w^5} & \quad |\Delta n| = 2
%     \end{array}.
%     \right.
% \end{equation}
% \textcolor{red}{Given $f(w,-d)=-2\sqrt{2}\frac{d}{w}e^{-d^2/w^2}$,The Rabi frequency is thus}
% \begin{equation}
%     \Omega_{n,n'} = \Omega_0f(w,-d)\sqrt{\frac{n_>!}{n_<!}}\times
%     \left\{
%     \begin{array}{cc}
%     1 -\frac{2x_0^2(3dw^2-2d^3)}{dw^4}(n+\frac{1}{2}) & \quad \Delta n = 0\\
%     -\frac{x_0(w^2-2d^2)}{dw^2} & \quad |\Delta n| = 1\\
%     -\frac{x_0^2(3dw^2-2d^3)}{dw^4} & \quad |\Delta n| = 2
%     \end{array}.
%     \right.\label{eq:TEM10SidebandStrengths}
% \end{equation}
% \textcolor{red}{Note that the carrier Rabi frequency is negative here. Note also that I have copied the form of eq.~\ref{eq:SidebandStrengths}, but by doing so it is not obvious that the even orders disappear for $d=0$ --- it is hidden in $f(w,-d)$.}
\begin{equation}
    \fl\Omega_{n',n} = \Omega_0 \,\, 2\sqrt{2} \,e^{-d^2/w_0^2} \sqrt{\frac{n_>!}{n_<!}}\times
    \left\{
    \begin{array}{cc}
    -\frac{d}{w_0} \left( 1 - \frac{6 x_0^2}{w_0^2} ( 1 -\frac{2 d^2}{3 w_0^2} )(n+\frac{1}{2}) \right) & \quad \Delta n = 0\\
    \frac{x_0}{w_0} (1 - \frac{2 d^2}{w_0^2} ) & \quad |\Delta n| = 1\\
    \frac{d}{w_0} \frac{3 x_0^2}{w_0^2} (1 - \frac{2 d^2}{3 w_0^2})& \quad |\Delta n| = 2
    \end{array}.
    \right.\label{eq:TEM10SidebandStrengths}
\end{equation}
% \begin{equation}
%     \hspace{50pt}\fl\Omega_{n^\prime n}= \Omega_0 \, \sqrt{\frac{n_>!}{n_<!}} \times
%     \left\{
%     \begin{array}{cc}
%     0 & \quad \Delta n = 0\\
%     2\sqrt{2}x_0/w & \quad |\Delta n| = 1\\
%     0 & \quad |\Delta n| = 2
%     \end{array}.
%     \right.\label{eq:TEM10SidebandStrengths}
% \end{equation}
Since $f(w_0,x)$ is odd, when $d=0$ the carrier and all even order sidebands vanish to all orders in $x_0$ (an exact expression for $\Omega_{n^\prime,n}$ for $d\!=\!0$ and arbitrary order Hermite-Gaussian profile can be found in \ref{sec:rabi_exact}).
%Likewise, for a TEM$_{00}$ mode, Eq.~(\ref{eq:SidebandStrengths}) shows that the first order sidebands vanish for $d=0$.
This suggests that by switching between transverse spatial modes, the carrier or first sidebands can be suppressed as the application demands, which can be used to reduce undesired off-resonant effects. A related effect in the longitudinal direction has been explored for optical standing waves \cite{Cirac1992Laser,Wineland1998,James1998Quantum,Reimann2014,Urunuela2020}, but the motional coupling in that case is still constrained to be along the longitudinal direction.

The appearance of sidebands (i.e.\ motional coupling) from the transverse spatial profile of a laser beam can be understood semi-classically in the time domain by considering that the oscillatory motion of an ion into and out of a laser beam gives an intensity modulation that produces sidebands at this oscillation frequency, which can in turn drive motional-state-changing transitions.
%\textcolor{red}{we never really thought much about rastering a laser beam as an equivalent process}
Alternatively, one can consider the associated Bloch sphere. In a frame rotating at the qubit splitting, the Bloch vector precesses azimuthally at a frequency equal to the detuning, $\Delta$. 
With negligible ion motion, no significant population transfer occurs (assuming $\Omega \ll \Delta$). With ion oscillation comparable to the beam size, the Rabi frequency will be modulated at $\omega$ and $2\omega$, associated with the linear and quadratic parts of $f(\ww,x)$, respectively. When $\Delta=\omega$, or $2\omega$, the precession and intensity modulation are synchronized. The result is a Bloch vector that `spirals' up or down the Bloch sphere even for $\Omega\ll\Delta$ (see Supplemental Material).

\section{Experiment}
\begin{figure*}[!ht]
    \centering
    \includegraphics[width=0.95\linewidth]{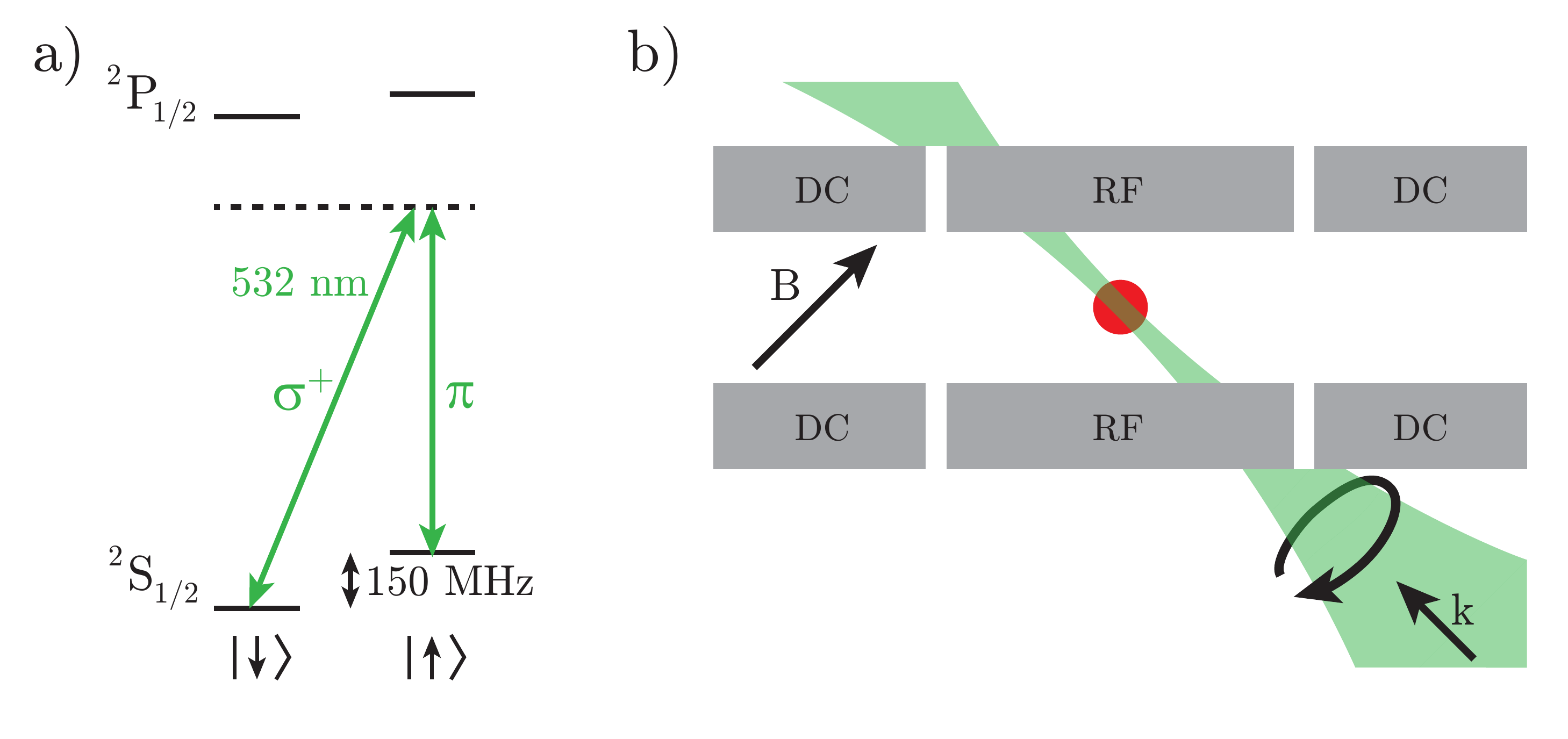}
    \caption{a) Structure of the ${}^{138}\mathrm{Ba}^+$ Zeeman qubit showing the laser field applied to drive stimulated Raman transitions b) Schematic of ion trap showing two of the four segmented rods and the single circularly polarised beam used to drive the Raman transitions.}
    \label{fig:exp_setup}
\end{figure*}

The analysis we have presented indicates that if motional coupling can be driven by the transverse profile of a laser beam, sidebands should appear even for a co-propagating stimulated Raman transition (for our setup, this gives a Lamb-Dicke parameter of $\eta \! \approx \!10^{-7}$).  The experiment we perform to observe these sidebands is shown schematically in Figure~\ref{fig:exp_setup}. Briefly, we trap a single laser-cooled $^{138}{\rm Ba}^+$ ion in a linear Paul trap made with four segmented cylindrical rods. The diagonal surface-to-surface distance between the rods is $2r_0 = 2~{\rm cm}$. RF voltages are applied to the central segments at a frequency of 1~MHz to produce a radial secular frequency $\omega_{\rm rad} \approx 2\pi\times100~{\rm kHz}$. The axial secular frequency is typically $\omega_{\rm ax} \approx 2\pi\times30~{\rm kHz}$.

We define a Zeeman qubit with the two electron spin states ($\kd$, $\ku$) of the $^2{\rm S}_{1/2}$ ground state manifold, which are split by 151.8~MHz by the application of a magnetic field of around 5.5~mT. Preparation of the qubit states is performed via optical pumping with circularly polarised light on the $^2{\rm S}_{1/2}$ $\leftrightarrow$ $^2{\rm P}_{1/2}$ transition. Readout of the qubit state is achieved via electron shelving; circularly polarised light at 455~nm selectively optically pumps one of the qubit states to the long lived ($\tau\approx 30~{\rm s}$) $^2{\rm D}_{5/2}$ manifold via the $^2{\rm P}_{3/2}$ manifold. Coherent transfer between the qubit states is driven by a far-detuned stimulated Raman transition via a mode-locked Nd:YVO$_4$ laser\footnote{Coherent Paladin SCAN 532-36000.}. The qubit splitting is close to twice the repetition rate of the laser such that different frequency components of the laser light can resonantly drive the qubit transition when the magnetic field tunes the qubit splitting into resonance. The use of a mode-locked laser for this type of manipulation has previously been demonstrated in work with hyperfine qubits \cite{Hayes2010}, but to our knowledge this is the first application to a Zeeman qubit. 

To observe sidebands, we direct a single (i.e.\ `co-propagating') circularly polarized beam at $45^{\circ}$ to the axis of the trap and at $90^{\circ}$ degrees to the quantization axis defined by the applied magnetic field (see figure~\ref{fig:exp_setup}). Even though none of the principal axes of the trap are prependicular to the laser beam, traditional (i.e.\ longitudinal) spin-motion coupling will be effectively absent for this co-propagating geometry, and the appearance of sidebands will be entirely due to transverse spin-motion coupling.  We perform Rabi spectroscopy on the Raman transition by measuring the spin flip probability while varying the applied magnetic field with a shim coil.

\begin{figure}[!ht]
    \centering
    \includegraphics[width=0.95\linewidth]{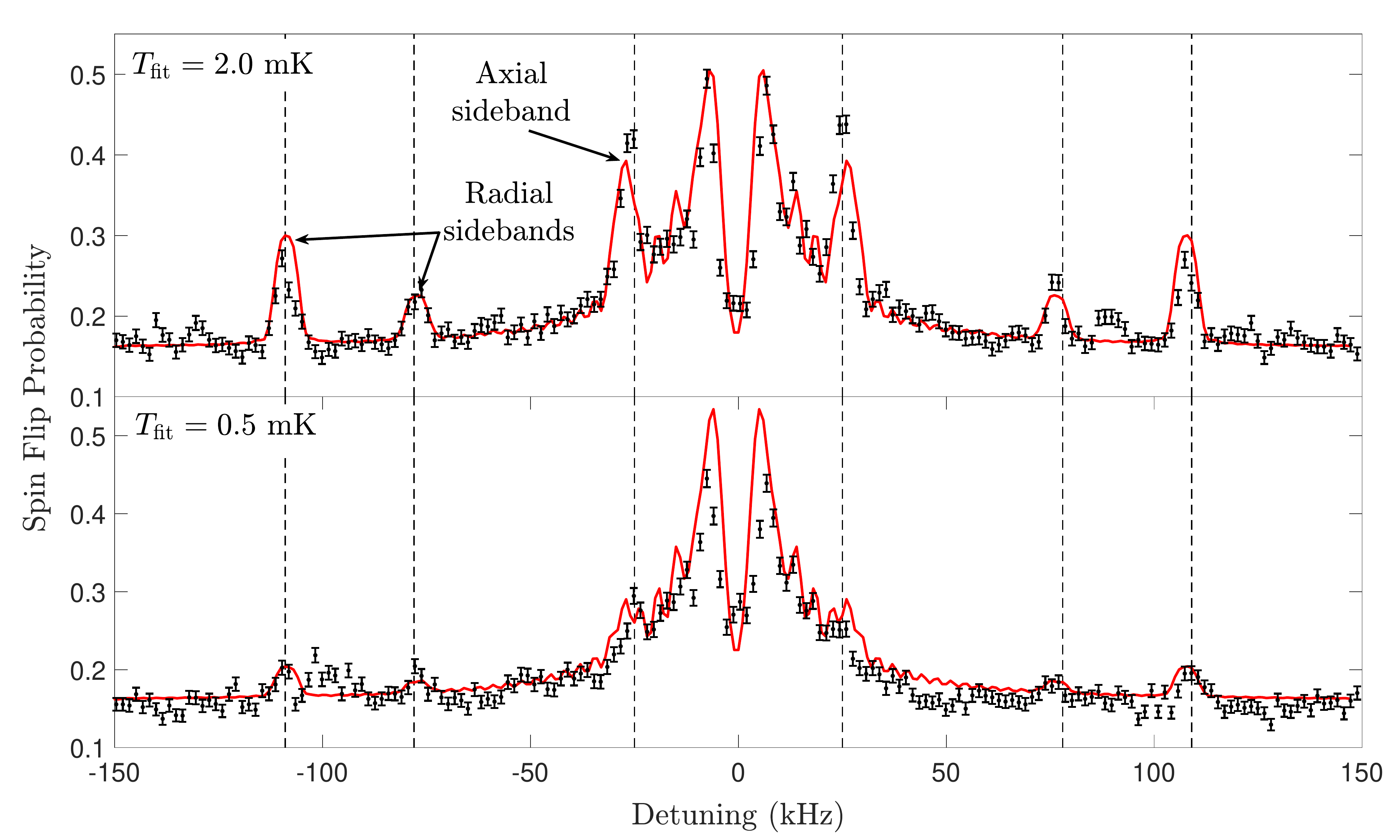}
    \caption{ Probability to measure $\kd$ as a function of detuning. Black: Experimental data, error bars represent the standard error on the mean from 2000 repetitions. Red: Fit based on theory --- see supplemental material for details. Vertical dashed lines indicate sideband features. Residual oscillations near zero detuning are due to a pulse area of $\approx8\pi$. The data in the two plots were taken under identical conditions except for a change in laser cooling efficiency.}
    \label{fig:fitted_spectrum}
\end{figure}

Figure~\ref{fig:fitted_spectrum} shows the probability of a stimulated Raman transition (with state preparation and measurement errors included) as a function of the detuning for two different temperatures. Motional sidebands associated with each of the three modes of motion (labelled with vertical dashed lines) are clearly visible --- two radial modes at $\omega/2 \pi \approx 110$~kHz and $80$~kHz, and the axial mode near $25$~kHz. For these data, the duration of the Raman pulse is equal to $8\pi$, resulting in the observed signal oscillations near zero detuning.

We calculate spectra via numerical solution of the Schrodinger equation with Rabi frequencies calculated as previously described and fit the result by varying the temperature, beam waist and beam position, assuming a Gaussian profile (see Supplemental Material for details). We also fit the coupling strength, $\Omega_0$, and secular frequencies, which are in good agreement with auxiliary measurements. In the upper plot of fig.~\ref{fig:fitted_spectrum}, the laser cooling was deliberately made inefficient by changing the detuning --- all other experimental parameters were kept the same. To account for magnetic field noise, which causes decoherence, we apply a boxcar average to the data of width 6~kHz.

\section{Discussion}
Spin-motion coupling due to the transverse electric-field profile presents an additional tool with which to manipulate trapped ions. However, it may also represent an additional source of infidelity in trapped ion quantum computers. In many cases, single-site addressability is required and achieved via tightly focused laser beams, which introduces spin-motion coupling for the transverse directions. 
% As an example, consider a $^{171}{\rm Yb}^+$ ion in a trapping potential with an axial secular frequency $\omega_{\rm ax} = 2\pi\times0.1~$MHz, corresponding to $z_0\approx17$~nm. Addressing it with a beam with a $1~\upmu$m waist gives $\etaeff\approx0.034$ for the second sidebands.  A displacement of $d=500~$nm would produce the same value of $\etaeff$ for the first sidebands.
\begin{figure}[!ht]
    \centering
    \includegraphics[width=0.4\linewidth]{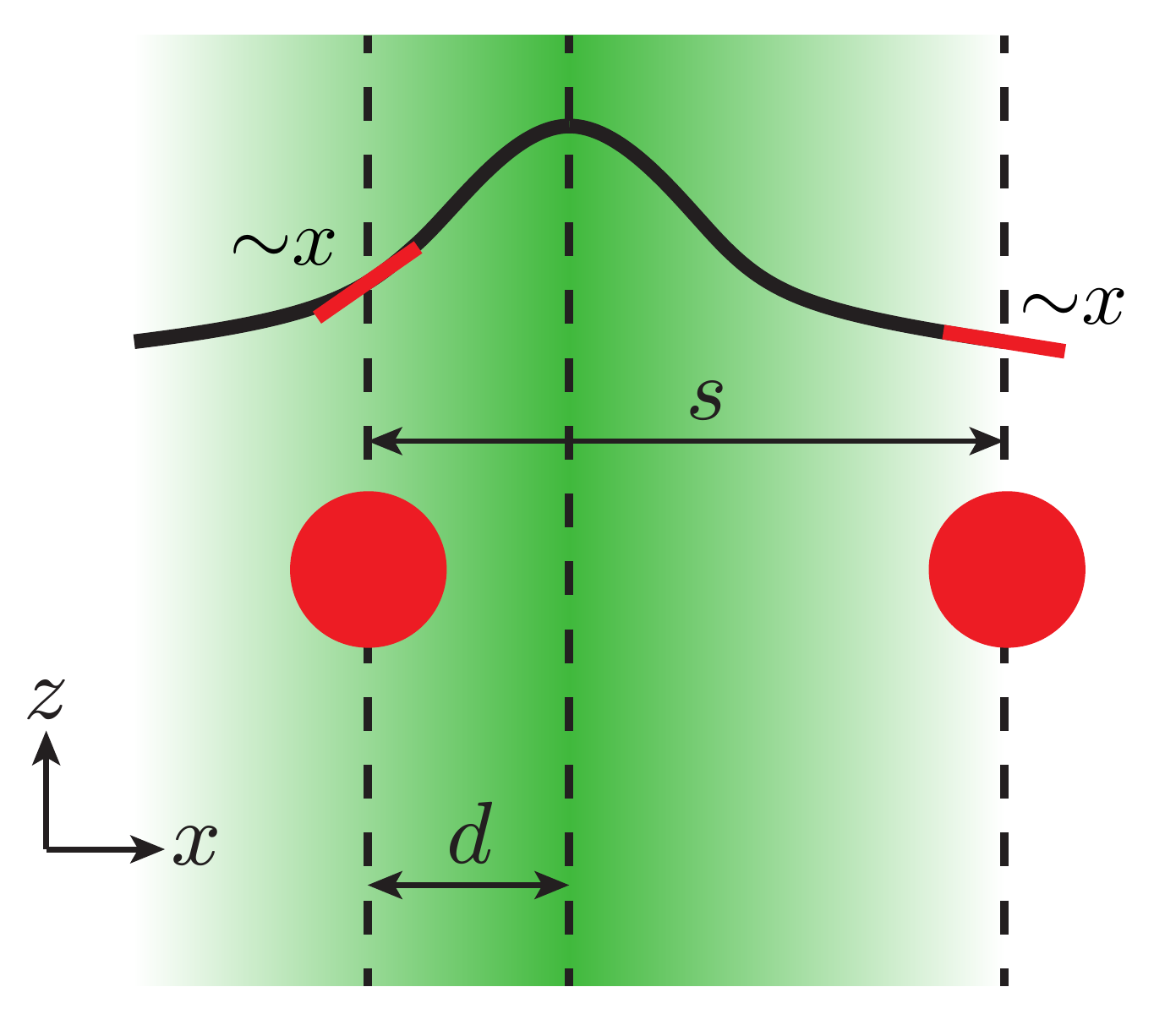}
    \caption{Laser incident on two ions, misaligned from a `target' ion (left) by $d$. The transverse spatial profile of the interaction strength is approximately linear for both the target ion and the neighbouring ion, which can produce spin-motion coupling along the $x$ direction.}
    \label{fig:schematic2}
\end{figure}
With reference to fig.~\ref{fig:schematic2}, we consider as an example the trapped ion quantum processor of Debnath \textit{et al.}~\cite{Debnath2016}. Individual $^{171}\mathrm{Yb}^+$ ions, spaced by $s\approx 5~\upmu$m in a trap with axial secular frequency $\omega/2\pi = 270$~kHz, are addressed by a pair of stimulated Raman beams. One of the beams provides a uniform intensity, while the other has waist $w\approx1.5~\upmu$m. The maximum quoted crosstalk of 4~\% (which we interpret here to mean the carrier Rabi frequency on a neighbouring ion is 4~\% of that of the target ion) could be produced by a Gaussian beam misalignment of $d\!\approx \! 1.8~\upmu\mathrm{m}$. For this value of misalignment, we can estimate the associated effective Lamb-Dicke parameters for transverse motional coupling from Eq.~(\ref{eq:SidebandStrengths}). For the target ion, we find for the first (second) sidebands that $\etaeff_{(1)} \approx 0.017$ ($\etaeff_{(2)} \approx 0.013$). Similarly, for the neighbouring ion we find $\etaeff_{(1)} \approx 0.030$ ($\etaeff_{(2)} \approx 0.028$). At the Doppler limit for $\mathrm{Yb}^+$, the mean axial phonon occupation would be $\bar{n}\approx 36$.  Since the resonant sideband Rabi frequencies are set by $\frac{\tilde{\eta}^p_{(p)}}{p!}\sqrt{\frac{n_>!}{n_<!}}$, transverse profile driven spin-motion coupling in the axial direction could lead to significant residual entanglement or other complications in this or other similar linear ion trap processors.

\ack{This work was supported by the Office of Naval Research (award N000141712256) and the Defense Advanced Research Projects Agency (award D18AP00067).}

\section*{References}
\bibliographystyle{iopart-num}
\bibliography{bibl.bib}

\appendix
\section{Exact Expressions for Transverse Rabi Frequency}
\label{sec:rabi_exact}
While the use of a Taylor series in eqs.~(\ref{eq:SidebandStrengths}) and (\ref{eq:TEM10SidebandStrengths}) provides intuition about how coupling between motional states depends on the transverse profile, exact analytic expressions for the Rabi frequency exist for the case $d=0$ (no beam misalignment). Below, we provide these expressions, which are based on integral identities of Hermite-Gaussian functions given in Refs.~\cite{Erdelyi} and \cite{Bailey1948Some}.

For the case of a TEM$_{00}$ mode, as long as $n^\prime + n$ is even, we have
\begin{eqnarray}
    \frac{\Omega_{n^\prime,n}}{\Omega_0} &=& \bra{n^\prime} e^{-2x^2/\sww^2} \ket{n}  \nonumber \\  &=&\frac{1}{\sqrt{2 \pi \, n^\prime ! \,n!}} \,a^{-n^\prime - n - 1} \left(1-2a^2 \right)^{\frac{n^\prime + n}{2}} \Gamma \! \left[ \frac{n^\prime + n + 1}{2}  \right]\nonumber\\
    &&\times{}_2F_1 \!\! \left[ -n, -n^\prime; \frac{1 - n^\prime -n}{2}; \frac{a^2}{2 a^2 - 1} \right] \label{eq:CenteredGaussian}
\end{eqnarray}
with
\begin{equation}
    a \equiv \sqrt{\frac{\ww^2 + 4 x_0^2}{2 \ww^2}}.
\end{equation}
Here, $\Gamma[x]$ is the Gamma function and ${}_2F_1[a,b;c;z]$ is the ordinary hypergeometric function.  As in Eq.~(\ref{eq:SidebandStrengths}), $\ww\!=\! \sqrt{2}w_0$ for single-photon transisions and $\ww\! = \!w_0$ for stimulated Raman transitions with both fields in the same TEM$_{00}$ mode.

For the case of a stimulated Raman transition with one arm having a TEM$_{00}$ mode and one arm having a TEM$_{p0}$ mode (both with the same waist $w_0$), as long as $n^\prime + n$ has the same parity as $p$, we have
\begin{eqnarray}
    \frac{\Omega_{n^\prime,n,p}}{\Omega_0} & = & \bra{n^\prime} H_p ( \sqrt{2}x/w_0 ) e^{-2 x^2/w_0^2} \ket{n}  \nonumber \\
    &=&  \frac{(-1)^{M-p}\,\, 2^{M+\frac{p}{2}}}{\sqrt{ \pi \,n^\prime !\, n! }} \,\, b^{n^\prime + n}\, a^{p+1} \,\, \Gamma \! \left[ M+ \frac{1}{2}\right]\,\nonumber\\
    &&\times {}_2F_1 \!\! \left[ -n, -n^\prime; \,\,\frac{1}{2} - M;\,\, \frac{1}{2 b^2} \right] \label{eq:GaussianAndTEMp0}
\end{eqnarray}
with
\begin{eqnarray}
a & \equiv & \frac{w_0}{\sqrt{w_0^2 + 4 x_0^2}}, \\
b & \equiv & \frac{2 x_0}{\sqrt{w_0^2 + 4 x_0^2}}, \\
M & \equiv & \frac{n^\prime + n + p}{2}.
\end{eqnarray}
This reduces to Eq.~(\ref{eq:CenteredGaussian}) for $p=0$ and $w_0 \equiv \ww$.

For the case of a single-photon transition driven by a TEM$_{p0}$ mode, or a Raman transition driven by a combination of a uniform field and a TEM$_{p0}$ mode, (again assuming $n^\prime + n$ and $p$ have the same parity) we have
\begin{eqnarray}
    \frac{\Omega_{n^\prime,n,p}}{\Omega_0}  & = & \bra{n^\prime} H_p ( \sqrt{2}x/w_0 ) e^{- x^2/w_0^2} \ket{n} \nonumber \\
    &=&  \frac{(-1)^p\,\, 2^{M+\frac{p}{2}}}{\sqrt{ \pi \,n^\prime !\, n! }} \,\,a \,\,(a^2-1)^{\frac{n^\prime + n}{2}} (b^2-1)^\frac{p}{2}\,\,  \Gamma \! \left[ M+ \frac{1}{2}\right] \nonumber \\
    &&\times \sum_{t=0}^{\mathrm{Min}(n^\prime,n)} \!\! \Bigg( \frac{(-n^\prime)_t (-n)_t}{t! \, \left(\frac{1}{2} - M \right)_t}\,\, 2^{-t} \, (1-a^2)^{-t}  \nonumber \\
    && \hspace{55pt}\times {}_2F_1 \! \left[2t - n^\prime -n, -p; \,\,\frac{1}{2} - M + t;\,\, z \right] \Bigg) \label{eq:TEMp0}
\end{eqnarray}
with $M$ as above, but
\begin{eqnarray}
a & \equiv & \frac{w_0}{\sqrt{w_0^2 +  2x_0^2}}, \\
b & \equiv & \frac{2 x_0}{\sqrt{w_0^2 + 2 x_0^2}}, \\
z & \equiv & \frac{1}{2} - \frac{ a b}{2 \sqrt{(a^2-1)(b^2-1)}},
\end{eqnarray}
and $(n)_m$ is the Pochhammer rising factorial.  Since the sum in (\ref{eq:TEMp0}) is finite, this expression can be evaluated to produce an extact result.  Similar to Eq.~(\ref{eq:GaussianAndTEMp0}), Eq.~(\ref{eq:TEMp0}) reduces to Eq.~(\ref{eq:CenteredGaussian}) for $p=0$ and $w_0 \equiv \ww/\sqrt{2}$.

\clearpage

\section*{\hspace{130pt}\Large Supplemental Material}
\section*{Simulated Spectra}
\begin{figure}[!ht]
    \centering
    \includegraphics[width=0.9\linewidth]{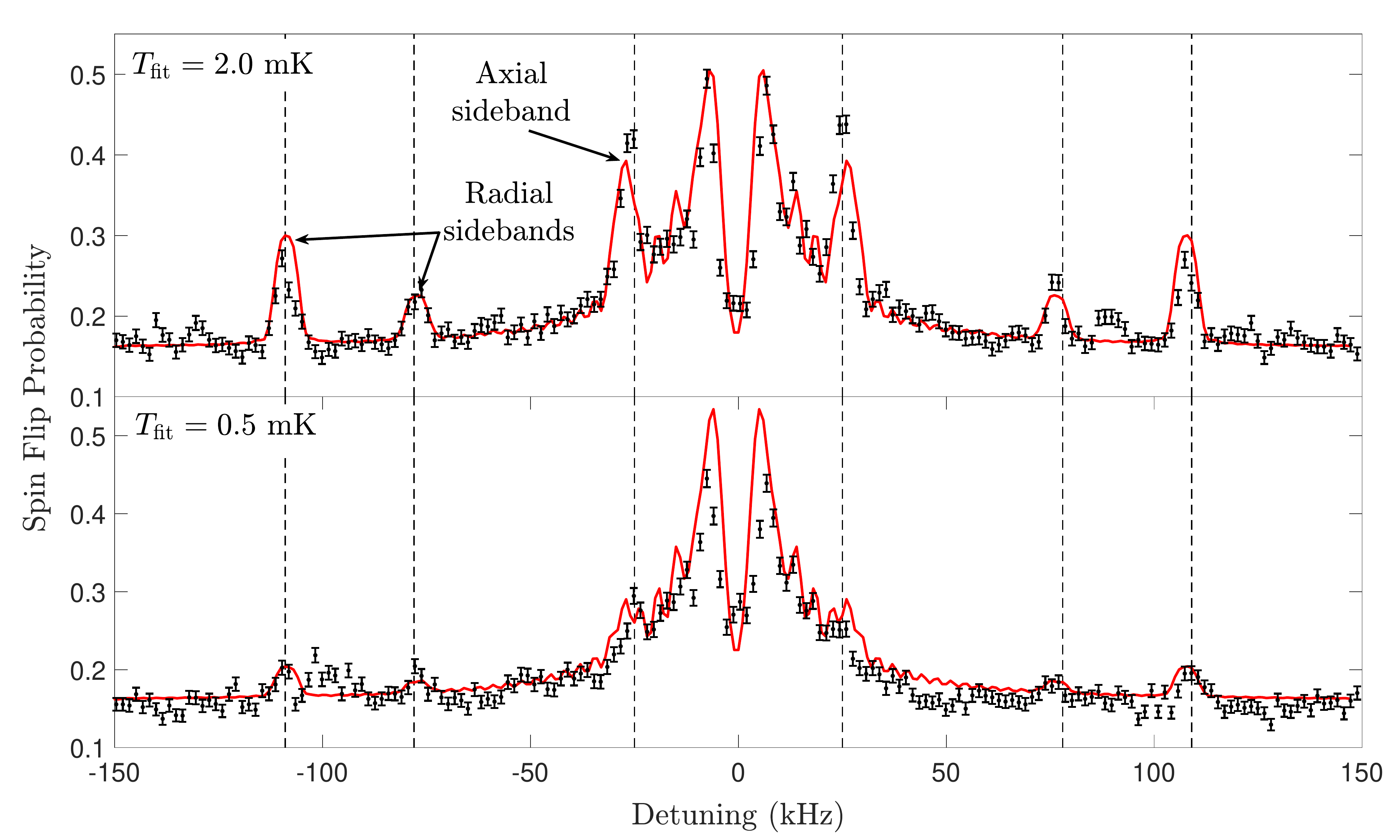}
    \caption{Probability to measure $\ku$ as a function of detuning. Black: Experimental data, error bars represent the standard error on the mean from 2000 repetitions. Red: Fit based on theory --- see main text for details. Vertical dashed lines indicate sideband features. Residual oscillations near zero detuning are due to the pulse area being significantly greater than $\pi$.}
    \label{fig:fitted_spectrum_supp}
\end{figure}
The figure above reproduces the spectra from the main paper. Black points represent measured data with error bars representing statistical uncertainty from 2000 repetitions. The red lines are simulated spectra obtained from a numerical solution of the Schrodinger equation. Thermal averaging is incorporated by averaging the result over a Boltzmann distribution of the initial motional states, $n_x,n_y,n_z$. The basis set includes all motional states with up to a total two quanta of motional excitation, i.e. $|(n'_x-n_x)|+|(n'_y-n_y)|+|(n'_z-n_z)|\le2$. All Rabi frequencies associated with these states are calculated according to chosen laser beam and trap parameters, up to and including all second-order sidebands.

In our system, the accuracy of the state measurement is limited due to off-resonant excitation during the shelving process. We model this by assigning a fidelity $f$ describing the probability of  measuring the correct qubit state. For a given probability of being in a particular state, $P$, the probability of measuring that state is then
\begin{eqnarray}
    P_{\rm meas} &=& P(2f-1)+1-f.
\end{eqnarray}
A value for $f$ can be estimated via numerical evaluation of the rate equations associated with the readout process, but this value is quite sensitive to drifts in laser power or polarization. As such, we estimate $f$ from the acquired spectra instead.

The following parameters were used to produce the fits in fig.~\ref{fig:fitted_spectrum}:
\begin{itemize}
    \item Carrier Rabi frequency at ion: $\Omega = 2\pi\times13.6$~kHz
    \item Raman pulse duration: $300~\upmu$s, $\Omega t \approx 8\pi$
    \item Beam waist: $15~\upmu$m along $x$, $14~\upmu$m along $y$
    \item Beam misalignment: $4~\upmu$m along $x$, $6~\upmu$m along $y$
    \item Secular frequencies: $\omega = 2\pi\times$109, 78, 30~kHz
    \item Detection fidelity: 0.84
\end{itemize}
Note that the $x$ and $y$ coordinates here are defined with a $z$ axis along the propagation direction of the beam (as opposed to according to the trap principal axes). The only difference in fit parameters for the two spectra shown in fig.~\ref{fig:fitted_spectrum} is the temperatures, which are as labelled. As stated in the main paper, a boxcar average of width 6~kHz is applied to the calculated spectrum to account for the presence of magnetic field noise in our system.

\clearpage

\section*{Semi-Classical Picture}
\begin{figure}[!ht]
    \centering
    \includegraphics[trim=200 100 180 100,clip,width=0.4\linewidth]{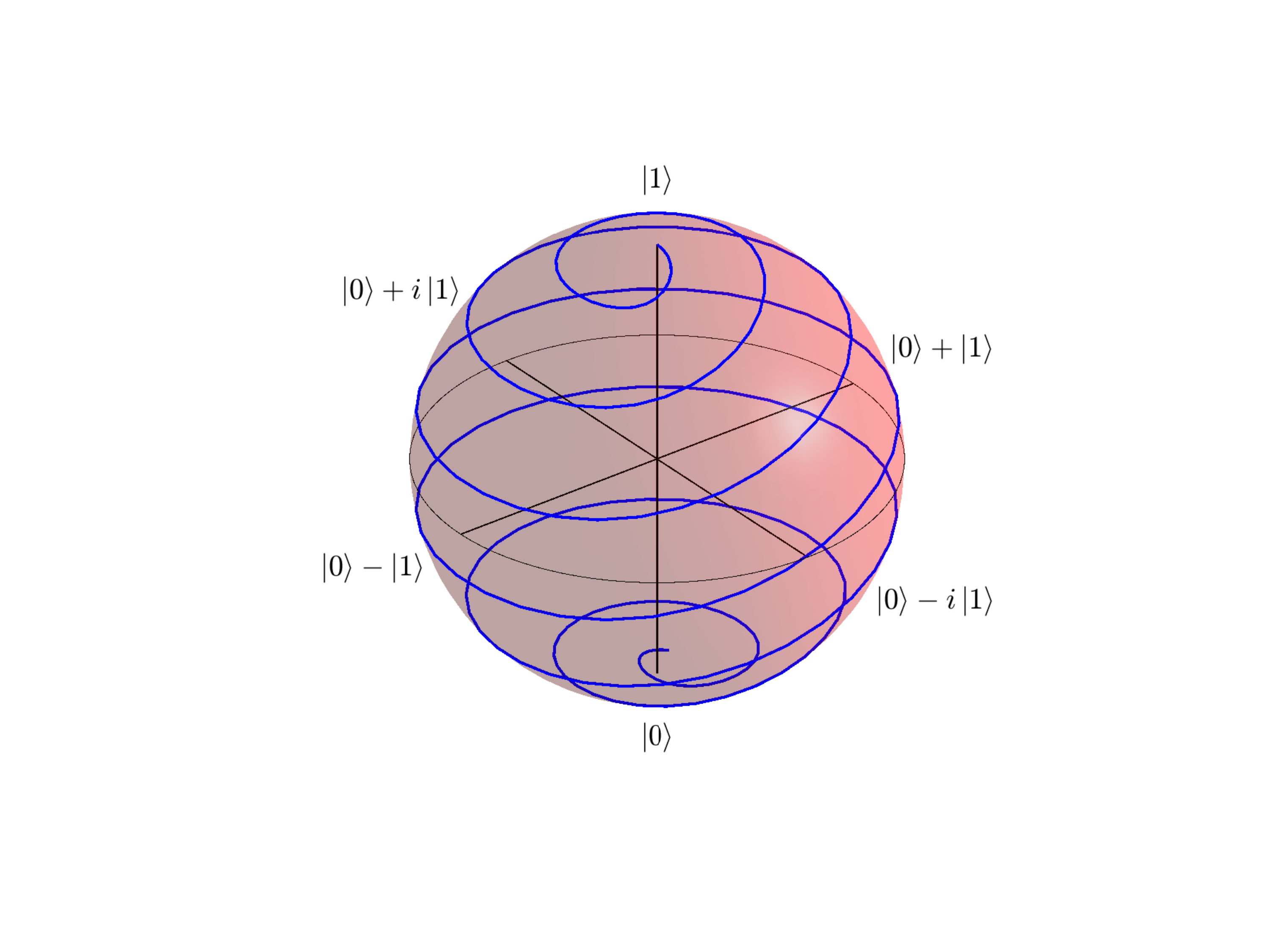}\hspace{60pt}
    \includegraphics[trim=200 100 180 100,clip,width=0.4\linewidth]{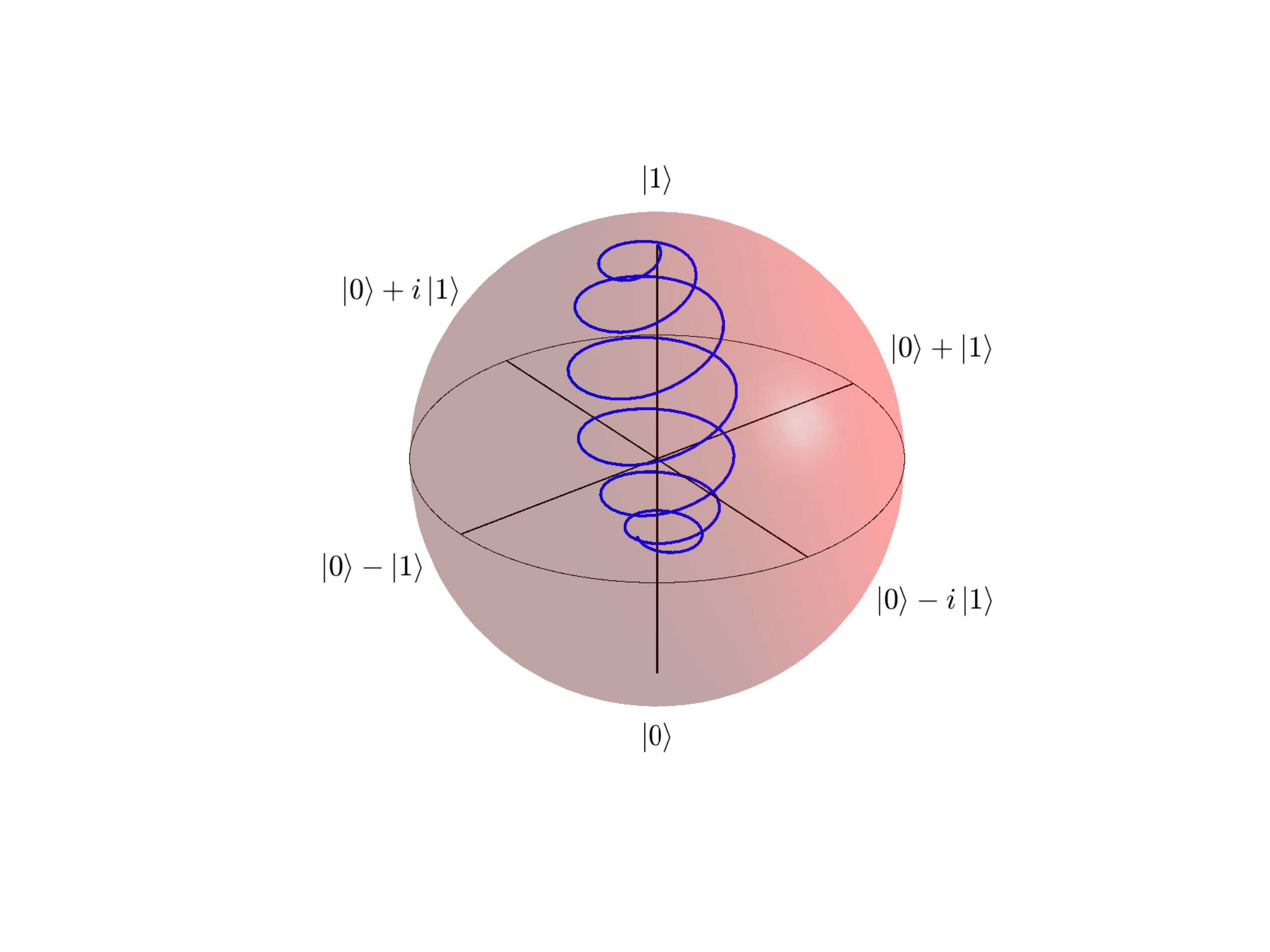}
    \caption{Plot of Bloch vector evolution for particles oscillating in a harmonic potential whilst addressed with an off-resonant laser beam with a Gaussian profile. Left: Single particle. Right: Average Bloch vector for 10 particles with different amplitudes and phases of motion.}
    \label{fig:blochtrails}
\end{figure}

The driving of motional sidebands can also be understood by considering the classical trajectory of an ion in a harmonic potential and the resulting evolution of the Bloch vector. In a frame rotating at the transition frequency, with an applied field detuned by $\Delta$, the motion of the Bloch vector is derived from two contributions: precession about the $z$ axis at a frequency $\Delta$ and rotation about the $x$ ($\ket{0}\pm\ket{1}$) axis at the instantaneous Rabi frequency, $\Omega(t)$. For a constant Rabi frequency which is less than the detuning, little population transfer occurs --- starting at $\ket{1}$, the Bloch vector is tipped towards $\ket{0}$ and then back towards $\ket{1}$ during each precession period, $1/\Delta$. Incorporating motion in the trap produces a modulated Rabi frequency. For a beam aligned with the trap center we have
\begin{eqnarray}
    \Omega(t) &=& \Omega_0\exp\left(-2x_0^2\sin^2(\omega t+\phi)/w^2\right)
\end{eqnarray}
where $x_0$ is the amplitude of motion, $\omega$ is the secular frequency, $\phi$ is the phase of the motional oscillation and $w$ is the beam waist. If the detuning is equal to the modulation frequency ($2\omega$ in this case), the Bloch vector will be preferentially rotate faster towards $\ket{0}$ and slower towards $\ket{1}$. This is illustrated in the left-hand plot of Fig.~\ref{fig:blochtrails}.

The result is that population transfer occurs on a motional sideband, without the need for counter-propagating beams. This behaviour persists when we consider a thermal ensemble which provides a distribution of $x_0$ and $\phi$. When we average over the thermal distribution, the coherence between the $\ket{0}$ and $\ket{1}$ states is lost. If we consider an ensemble Bloch vector which is the average of individual particle Bloch vectors, the loss of coherence causes this ensemble vector to lie along $\pm z$. This behaviour is shown in the right-hand plot of Fig.~\ref{fig:blochtrails}. Incorporating the thermal ensemble also reduces the degree of population transfer somewhat.

\end{document}